\documentclass[letterpaper, 10 pt, conference]{ieeeconf}  

\IEEEoverridecommandlockouts                              

\usepackage{epsfig} 
\usepackage{amsmath} 
\usepackage{amssymb}  

\usepackage{epstopdf}
\title{\LARGE \bf
Mechanically Powered Motion Imaging Phantoms: Proof of Concept
}

\author{Alberto~Gomez$^{1}$ \and Cornelia~Schmitz$^{1}$ \and Markus~Henningsson$^{2}$ \and James~Housden$^{1}$ \and Yohan~Noh$^{1}$ \and Veronika~A.~Zimmer$^{1}$ \and James~R.~Clough$^{1}$ \and Ilkay~Oksuz$^{1}$ \and Nicolas~Toussaint$^{1}$ \and Andrew~P.~King$^{1}$ \and Julia~A.~Schnabel$^{1}$,~\IEEEmembership{Senior~Member,~IEEE}%
\thanks{*This work was supported by the Wellcome Trust IEH Award [102431]. This work was supported by the Wellcome/EPSRC Centre for Medical Engineering at King’s College London [WT 203148/Z/16/Z]. The research was funded/supported by the National Institute for Health Research (NIHR) comprehensive Biomedical Research Centre awarded to Guy's and St Thomas' NHS Foundation Trust and King's College London. The views expressed are those of the author(s) and not necessarily those of the NHS, the NIHR or the Department of Health.}%
\thanks{$^{1}$A. Gomez, C. Schmitz, J. Housden, Y.~Noh, V.~A.~Zimmer, J.~Clough, I. Oksuz N.~Toussaint, A.~P.~King and J.~A.~Schnabel are with with the School of Biomedical Engineering and Imaging Sciences, King's College London, United Kingdom  
{(\tt\small alberto.gomez@kcl.ac.uk})}%
\thanks{$^{2}$M. Henningsson is with the Division of Cardiovascular Medicine, Department of Medical and Health Sciences, Link{\"o}ping University, Sweden}
}

\begin{document}

\maketitle
\thispagestyle{empty}
\pagestyle{empty}

\begin{abstract}
Motion imaging phantoms are expensive, bulky and difficult to transport and set-up. The purpose of this paper is to demonstrate a simple approach to the design of multi-modality motion imaging phantoms that use mechanically stored energy to produce motion.

We propose two phantom designs that use mainsprings and elastic bands to store energy. A rectangular piece was attached to an axle at the end of the transmission chain of each phantom, and underwent a rotary motion upon release of the mechanical motor. The phantoms were imaged with MRI and US, and the image sequences were embedded in a 1D non linear manifold (Laplacian Eigenmap) and the spectrogram of the embedding was used to derive the angular velocity over time.
The derived velocities were consistent and reproducible within a small error. The proposed motion phantom concept showed great potential for the construction of simple and affordable motion phantoms.
\end{abstract}
%

\section{INTRODUCTION}

Motion phantoms are a useful tool to validate and develop new techniques in medical imaging. In particular, cardiac and respiratory motion are an important source of imaging artefacts, especially for relatively slow imaging modalities such as magnetic resonance imaging (MRI) \cite{henningsson2012whole}, \cite{scott2009motion}, Computed Tomography (CT) \cite{ionasec2010patient} or Positron Emission Tomography (PET) \cite{huang2014mr}. Moreover, motion modelling, tracking and quantification are active areas of research,  
and require controlled motion data for validation, ideally from multiple sources (modalities). 

Most published work acknowledges motion phantoms as an essential tool for controlled motion experiments \cite{cloonan20143d}. Effectively, two types of motion imaging phantoms are used: electrically powered phantoms, and phantoms powered by pressurised air. In both cases, the phantoms rely on a pump which must be located outside the scanner. 
In electrically powered phantoms (the most common type), the pump has an electric motor that is unsafe in the proximity of the MRI scanner, hence must be placed outside the 5 Gauss line, and connected to the phantom through a non-ferromagnetic transmission system \cite{gaddum2014starling}. As a result, electrically powered phantoms require cumbersome set-ups and are relatively large and therefore difficult to transport. In the recently introduced air powered phantoms \cite{yue2018mri}, the phantom is connected to a compressed air source, and the air pressure drives the moving parts through a pipe system. Compressed air sources are relatively common in hospital settings, however they are not widely available elsewhere. Both types of phantoms are typically expensive, bulky and large, rendering them inconvenient for use with ultrasound systems, were physical contact with the transducer (or easy placement of the transducer within a fluid medium) is required. For this reason, as well as in the pursue for simplicity and portability, Grice et al. \cite{grice2016new} proposed a flow phantom for Doppler ultrasound which used gravity to move blood-mimicking fluid through a tube, by having two connected reservoirs placed at different heights.  Basically, energy is stored as potential energy by manually sending fluid to the upper reservoir, and released as kinetic energy of the moving fluid by opening a valve. 
However, this design does not allow any other type of motion, for example intra-cardiac flows \cite{Gomez15}.

We propose to use a novel type of motion phantoms which do not require an external power source and are compatible with multiple imaging systems including MRI and ultrasound (US): mechanical energy storing motion phantoms. This new class of phantoms is flexible enough to simulate a wide variety of physiological motion patterns, including fluid flow and tissue motion. More specifically, we propose two self-contained phantoms: an in-house 3D printed model with a wind-up spring and a Lego built model with elastic bands. 


Most methods for image-based motion quantification (e.g. registration, tracking) require tuning modality-specific parameters, therefore introducing a potential source of discrepancy when comparing across modalities. In this paper, we propose a novel, modality independent method to quantify periodic motion, and discuss potential extensions to other types of motion.

The contributions of this paper are threefold: first, we propose a novel type of motion mechanism to make motion imaging phantoms. Second, we demonstrate two examples of rotary motion phantoms under MRI and US imaging. Finally, we propose a image analysis pipeline to estimate the rotary velocity over time from the acquired images.

\section{MATERIALS AND METHODS}
\label{sec:materials-methods}

\subsection{Design Considerations}
\label{sec:methods:design-considerations}
In order to make a multi-modality motion imaging phantom usable for medical imaging research, the design must:
\begin{enumerate}
    \item Not have any electric or ferromagnetic parts (for MRI safety and minimum artefacts in CT and US).
    \item Be water resistant (many experiments and particularly US imaging are likely to involve a water tank).
    \item Produce velocities in the physiological range, e.g. $[1,10]$ cm/s for tissue, and $[10, 100]$ cm/s for blood.
    \item Last for longer than one physiological cycle, e.g. $>1s$ for cardiac motion, and $>1m$ for respiratory motion.
    \item Be customisable to the application of interest, easy to transport and set-up, and cost-effective.
\end{enumerate}

In this section we propose designs that meet the requirements listed above using mechanically stored energy. 

\subsection{Mechanically Powered Motion Phantom}

\begin{figure}[thpb]
\begin{minipage}[b]{\linewidth}
  \centering
  \centerline{\includegraphics[width=\linewidth]{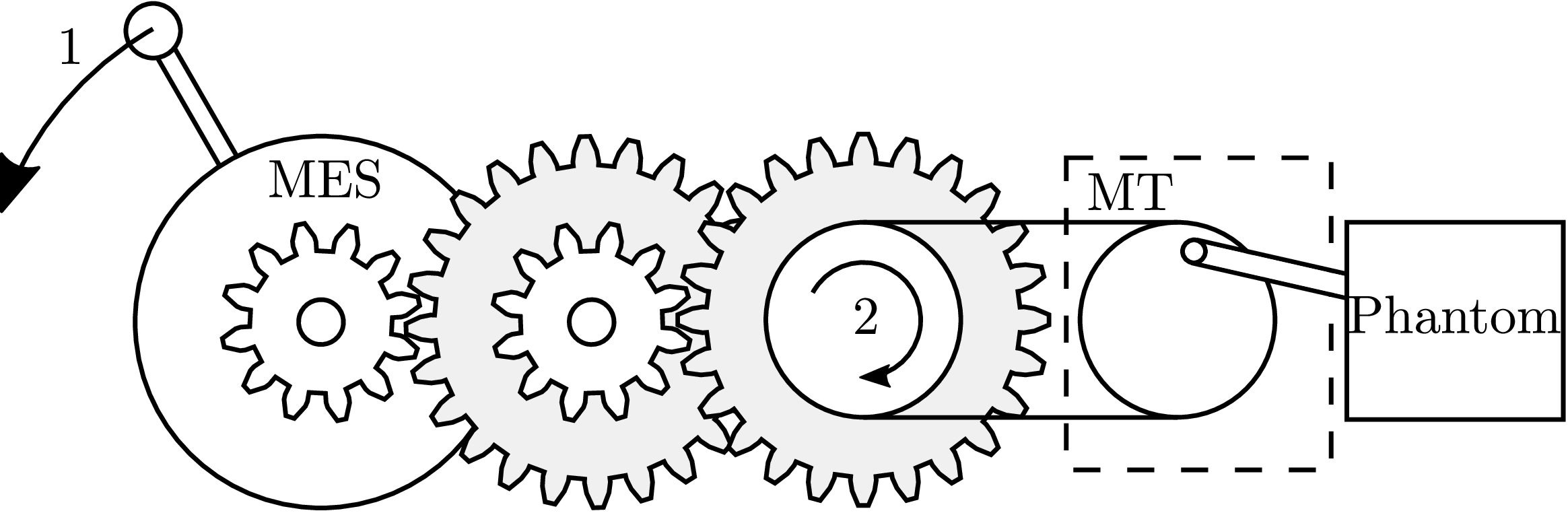}}
  \centerline{(a) Motion phantom concept\label{fig:transmission_chain}}\medskip
\end{minipage}

\begin{minipage}[b]{.48\linewidth}
  \centering
  \centerline{\includegraphics[width=\linewidth]{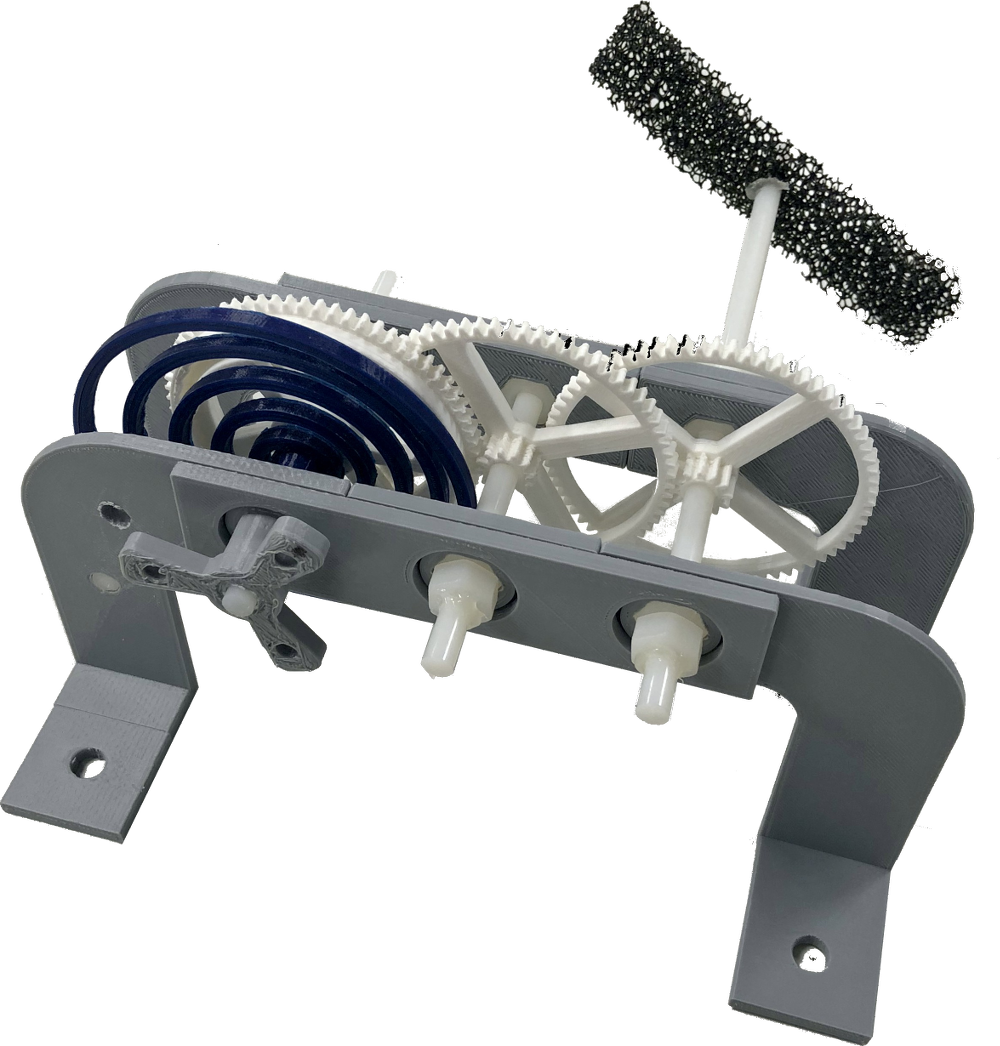}}
  \centerline{(b) 3D printed phantom\label{fig:phantom_3dp}}\medskip
\end{minipage}
\begin{minipage}[b]{.48\linewidth}
  \centering
  \centerline{\includegraphics[width=\linewidth]{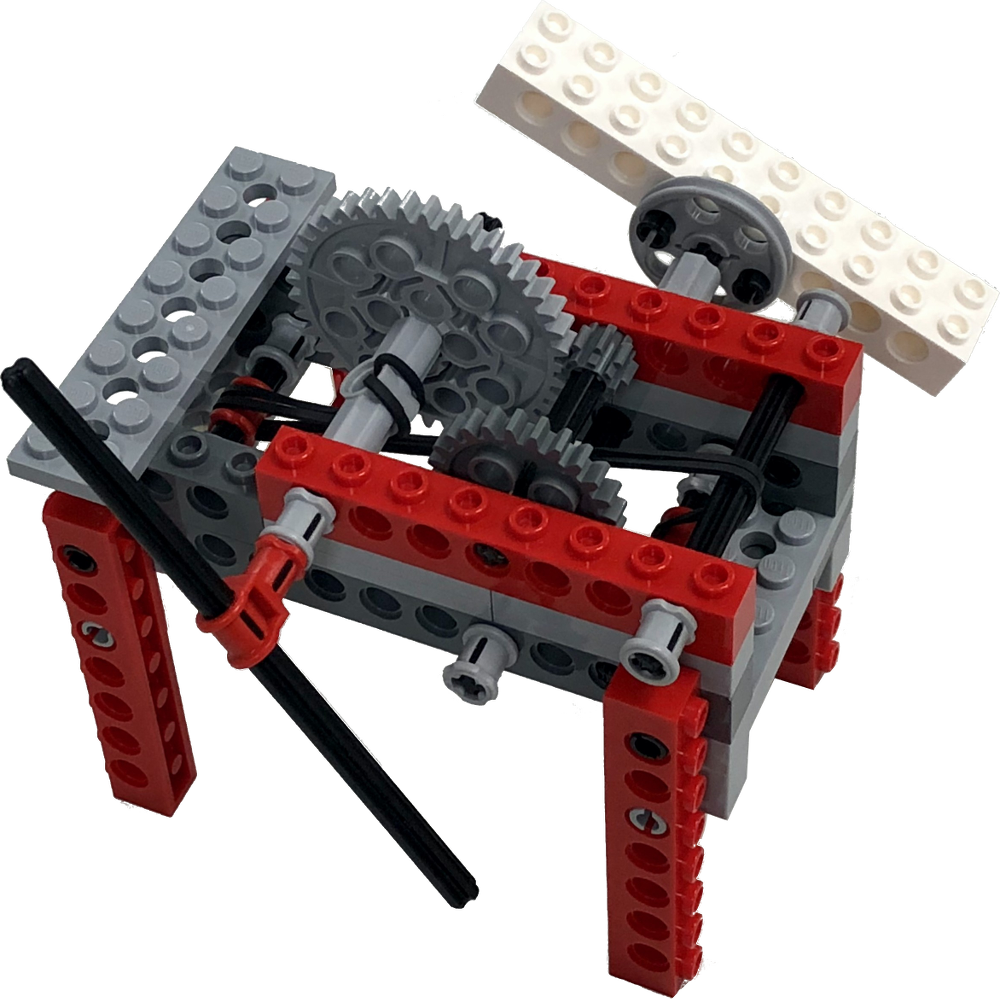}}
  \centerline{(c) Lego phantom\label{fig:phantom_lego}}\medskip
\end{minipage}

\caption{Phantom design concept. (a) A Mechanical Energy Storage (MES) subsystem  is loaded by turning a lever (1). When the MES is released a transmission chain propagates the power to a end axle (2) which is attached to a Motion Transformation (MT) subsystem that converts the rotatory motion into the desired motion. (b),(c) Our two implementations of the proposed concept.
}
\label{fig:phantoms}
\end{figure}

The proposed motion phantom concept is illustrated in Fig. \ref{fig:phantoms} (a). We propose two simple designs: a custom 3D printed model (Fig. \ref{fig:phantoms} (b)), where the Mechanical Energy Storage (MES) subsystem is a spiral spring  and the phantom is made of a piece of reticulated foam; and a mechanism made with Lego (Fig. \ref{fig:phantoms} (c)), where the MES is an elastic band and the phantom is made of Lego bars. In both cases, the Motion Transformation (MT) subsystem is a geared transmission chain which propagates a rotary motion to the phantom.

\subsection{3D Printed Phantom}

We demonstrate the customised fabrication of motion phantoms by adapting an available wind-up mechanism \cite{thingiverse1}. We designed a new stand, knob, and mainspring. 
%
The mainspring is a spiral which comes out tangent to the central axle. This can be parameterised as follows:
\begin{equation}
    \begin{array}{cc}
    x(\theta) &= R_1-(R_0-R_1)\frac{\theta}{2\pi N} \cos(\theta) \\
    y(\theta) &= R_1-(R_0-R_1)\frac{\theta}{2\pi N} \sin(\theta) \\
    \end{array}
\end{equation}
where $R_0$ is the inner radius (axle radius), $R_1$ is the outer radius, and $N$ is the number of turns (pitch) of the mainspring at rest. The last gear of the transmission system was attached to a $11\times 2 \times 1.5$cm block of open-pore reticulated foam (Fig. \ref{fig:phantoms} (b)). 
The foam block rotated  around its centre when the loaded mainspring was released. The stand had four attachments in its base to fix it to the bottom of a plastic water tank for the imaging experiments. The prototype parts were printed in PLA using a Delta WASP 2040 Turbo2 printer.

\subsection{Lego Phantom}

We demonstrate the use of low-cost motion phantoms by proposing a Lego based design inspired by the Stillinger's \emph{Supercharged Speedster}~\cite{stillinger2008crazy}. In this design, energy is stored in an elastic band that is winded around the axle of the first gear of a transmission chain, the last gear is connected to a $10 \times 2 $ Lego Technic bar  (Fig. \ref{fig:phantoms} (c)), which rotates around its centre when the loaded band is released. As with the other phantom, four Lego blocks were glued to the bottom of a plastic water tank to attach the phantom for the experiments.

\subsection{Experiment Design}

Both phantoms were imaged with US and MRI. US imaging was carried out with a Philips EPIQ V7 and a X6-1 transducer which was statically held at a fixed position such that the 2D imaging plane contained the rotating bar of the phantom completely. US acquisition settings were tuned to achieve a frame rate of 20Hz at maximum width in 2D mode. 
MRI was carried out with a 1.5T Philips Ingenia scanner, using a 2D balanced steady-state free precession (bSSFP) sequence, with a temporal resolution of $50$ ms. 

\begin{figure}[thpb]
\centering
\includegraphics[width=\linewidth]{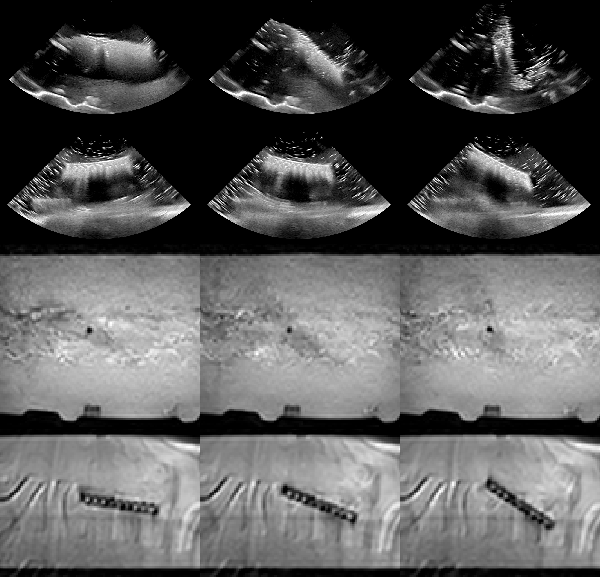}%
\caption{Example of images acquired at three times. 1st row: US, 3DP phantom. 2nd row: US, Lego phantom. 3rd row: MRI, 3DP phantom. 4th row: MRI, Lego phantom.}
\label{fig:images}
\end{figure}

For the experiments, a $32\times45\times25$cm plastic water tank was filled up with water and located on a workbench (US) or inside the scanner (MRI). The mechanism was wound up by the operator, to a fixed number of turns for each phantom. Imaging was started and immediately after the operator released the phantom, until rotation stopped completely. The same process was repeated 5 times for each modality and each phantom, totalling 20 acquired sequences (Fig. \ref{fig:images}).

\section{DATA ANALYSIS}
\label{sec:data-analysis}

Feature tracking and image registration techniques require the non-trivial set-up of a number of registration parameters (similarity measure, interpolation type, or optimisation method to cite a few) that are specific to each imaging modality. Hence, comparison across different modalities, which may require different parameters, is particularly challenging. In this paper, we propose to exploit the repetitive pattern in the appearance of the images of the rotating objects to quantify the time-varying angular velocity in a fully automatic, robust, and modality-independent way. 

\begin{figure}[thpb]
\centering
\includegraphics[width=\linewidth]{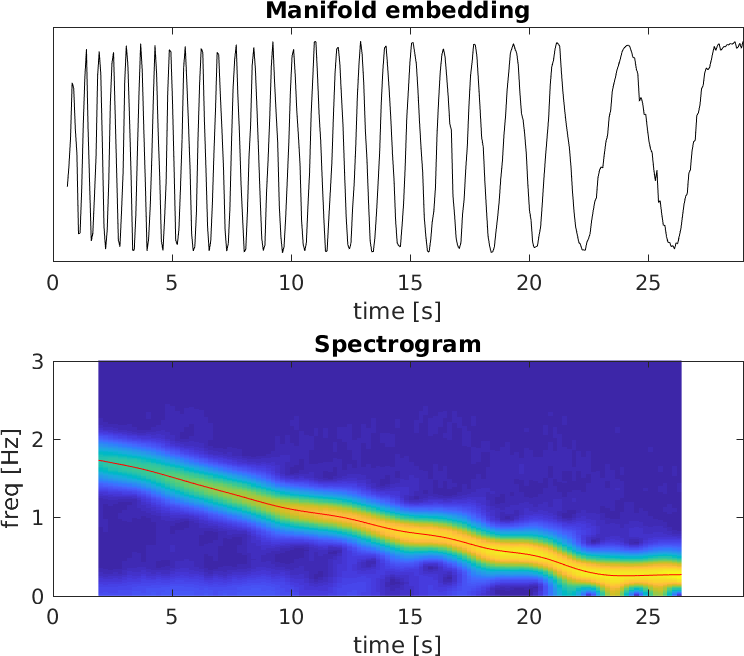}
\caption{Quantification of rotary motion using a spectrogram analysis of the manifold embedding of the imaging sequence. (Top) Laplacian Eigenmap embedding of one of the 3D printed phantom US sequences. (Bottom) Spectrogram of the above embedding, which when sampling rate is correctly accounted for yields the angular velocity over time.}
\label{fig:embedding-spectrogram}
\end{figure}

Because the in-plane rotation of a block can be described with one parameter at each frame (the angle), it follows that our acquired images can be embedded in a 1D manifold, and that the embedding will be representative of the angle. If this was true, the image sequence would be represented as a frequency-decreasing sinusoidal wave, and a time-frequency analysis of the embedding would yield the angular velocity of the block as a function of the time. Figure \ref{fig:embedding-spectrogram} (top) shows an example of such embedding using Laplacian Eigenmaps \cite{belkin2002laplacian}, confirming our hypothesis.

We carried out a spectrogram analysis (Fig. \ref{fig:embedding-spectrogram}) of the manifold embedding. The spectrogram $\mathcal{S}_\tau$ of a signal $r(t)$ is the power of the short-time Fourier transform (STFT), this is:
\begin{equation}
\mathcal{S}_\tau(r) = |STFT_\tau(r) |^2(\omega,t)
\end{equation}
where $\tau$ is a time interval to window the computation of the Fourier transforms, and from Nyquist-Shannon sampling theorem must be chosen bigger than twice the inverse of the lowest velocity to be measured (in our case, the lowest velocity before the mechanism stops is about 0.25 Hz). 

The energy is clustered over a curve (in yellow) that indicates the main frequency of the embedding. It is worth noting that this frequency is equal to half the rotation rate, because each cycle in the embedding is half a rotation --bar orientations 180 degrees apart are represented in the same region in the manifold.  
A spline was fitted to the spectrogram (weighted by the power). To remove unnecessary points, for each time sample only the $5\%$ higher power rows were used.


\section{RESULTS}
\label{sec:results}

\begin{figure}[thpb]

\begin{minipage}[b]{0.49\linewidth}
  \centering
  \centerline{\includegraphics[width=\linewidth]{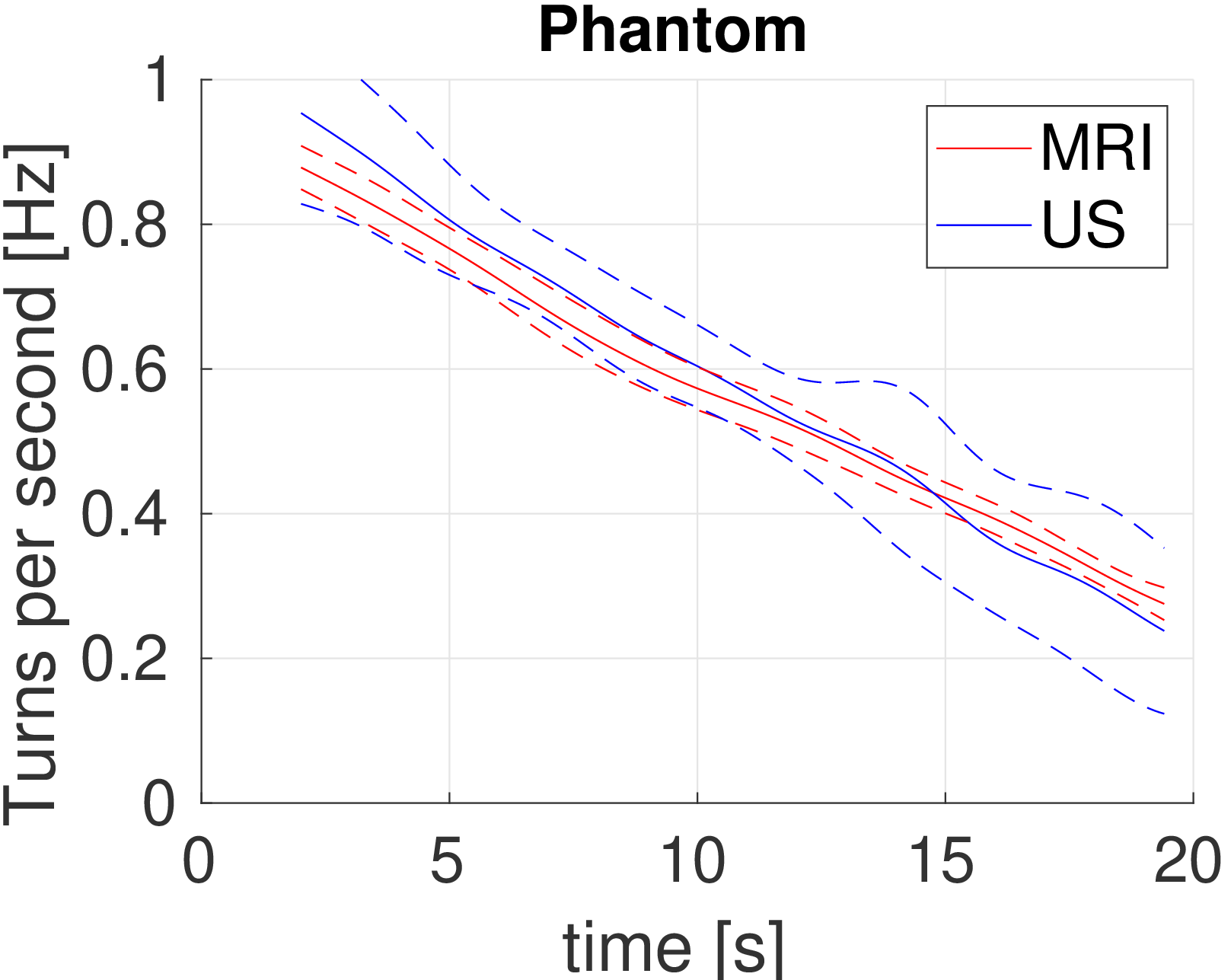}}
  \centerline{(a) 3D printed phantom\label{fig:vel-3dp}}\medskip
\end{minipage}
\begin{minipage}[b]{0.49\linewidth}
  \centering
  \centerline{\includegraphics[width=\linewidth]{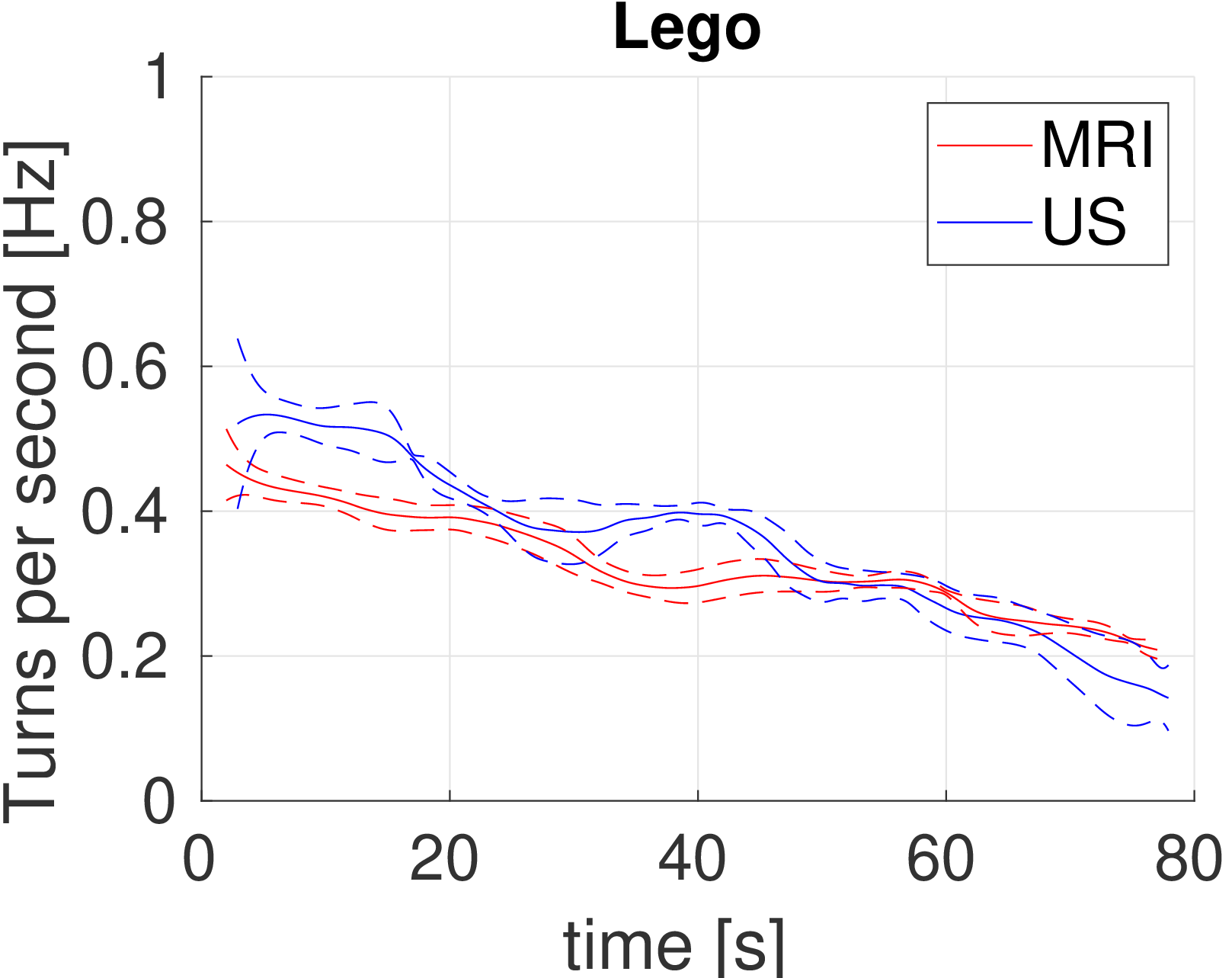}}
  \centerline{(b) Lego phantom\label{fig:vel-lego}}\medskip
\end{minipage}

\caption{Inter modality comparison of the angular velocities, showing MRI in blue and US in red. The solid lines indicate the average velocity over 5 repeated experiments, and the dashed lines indicate the standard deviation from the mean. }
\label{fig:vel-traces}
\end{figure}

A summary of the velocity consistency is given in Table \ref{tab:results}. For every phantom and imaging modality, the five velocity traces were compared over time, and gave the average angular velocity (represented as a solid line in Fig. \ref{fig:vel-traces}), and the velocity dispersion (the difference between the average velocity and each individual trace). The first two columns of table \ref{tab:results} show the root mean square (RMS) of this intra-modality dispersion. The third and fourth columns measure the RMS and the average $\pm$ standard deviation (respectively) over the inter-modality difference in average velocities for each phantom. The last column reports the maximum angular velocity (on average over all experiments).

\begin{table}[thpb]
\renewcommand{\arraystretch}{1.3}
\caption{Quantitative measurements of velocity for two phantoms under two imaging modalities, in Hz}
\label{tab:results}
\centering
\begin{tabular}{r|c|c||c|c || c}
\hline
& RMS & RMS & RMS &  av $\pm$ std & max\\
& MRI & US  &  MRI-US & MRI-US & vel.\\
\hline
\footnotesize{3DP} & $0.02$ & $0.08$ & $0.04$  & $-0.02\pm0.03$ & 0.96 \\
\footnotesize{LEGO}& $0.02$ & $0.03$ & $0.06$&  $-0.03\pm0.05$ & 0.55 \\
\hline
\end{tabular}
\end{table}

The MRI derived velocity is more reproducible, with a lower RMS value of approximately 2 to 3 \% of the maximum velociy (up to a 8\% for US, 3D printed phantom). 
A one-way ANOVA test between the MRI and the US derived velocities yield no statistically significant difference for the 3D printed phantom data ($p>0.1$). 
There was however a statistically significant difference between the average velocities derived for the Lego phantom ($p<0.01$) from MRI and US. The traces in Fig. \ref{fig:vel-lego} show a more unstable behaviour than the curves for the 3D printed phantom, and show differences in angular velocity especially during the first $45$ seconds of the experiment, after which the traces overlap. Velocities in the US experiment were higher than in MRI in all cases, by a 2\% to 3\% on average for both phantoms.

\section{DISCUSSION}
\label{sec:discussion}

Mechanically powered phantoms can fulfil all requirements from Sec. \ref{sec:methods:design-considerations}. Specifically, our two implementations of the proposed concept are made of plastic, hence  compatibility with all imaging modalities and water resistance is ensured. They run for approximately 20 seconds (the 3D printed model) and 80 seconds (the Lego model), which is enough to simulate multiple physiological cycles of cardiac and respiratory motion. 


Our results show that US provided a lower reproducibility for the quantification of the angular velocities,  likely due to poorer image quality compared to the MRI sequences, since the frame rate used for both modalities was very similar and sufficient for the rotary motion. Velocities derived from US were consistently higher than velocities derived from MRI for the same phantom. 
This could be explained by small differences in the experimental set-up, namely slightly different amount of water (therefore different pressure exerted on the rotating block). Also, deceleration in the Lego phantom is more irregular than the nearly linear deceleration observed in the 3D printed phantom. This may suggest that springs are a more stable energy storage system than elastic bands, although this needs to be further investigated. Both phantoms had a small footprint, are lightweight, and easy to set-up. Access to water is required to fill the bucket before the imaging procedure but otherwise the phantoms are self contained.

Future work should include experiments with other MES subsystems, (e.g. pressurised air, elevated weights), and mechanisms for steady energy release and pulsatile motion patterns. This will enable motion patters similar to those of interest within the human body, such as lung displacement over the respiratory cycle, cardiac motion, and fluid flow which could be achieved by using a propeller or a plunger attached in the motion transformation (MT) subsystem. Our work demonstrated that the proposed phantoms can be used to move solid tissue-like materials.  

A limitation of the current design is that, as the potential energy is released, it is transformed into kinetic energy until the mechanism stops. This effect prevents from producing a constant continuous motion. MT subsystems which can produce constant average motion by releasing the energy in bursts (e.g. the mechanisms in mechanical clocks) exist, but are out of the scope of this paper and are also left for future work. 

Our results demonstrate the utility of mechanical storage mechanisms to build reproducible, portable and cost effective multi-modality motion imaging phantoms. 


%


\end{document}